\documentclass[runningheads]{svmult}

\usepackage{physprbb}  

\newcommand{\greeksym}[1]{{\usefont{U}{psy}{m}{n}#1}}
\newcommand{\umu}{\mbox{\greeksym{m}}}

\begin{document}
\title*{A search for emission line galaxies at $z = 6.5$}
\toctitle{A search for emission line galaxies at $z = 6.5$}
%
%
\titlerunning{Emission line galaxies at $z = 6.5$}
%
\author{Jaron Kurk\inst{1}
\and Andrea Cimatti\inst{1}
\and Sperello di Serego Alighieri\inst{1}
\and Jo\"el Vernet\inst{1}
\and Emanuele Daddi\inst{2}}
\authorrunning{Jaron Kurk et al.}
%
%
\institute{
INAF -- Osservatorio Astrofisico di Arcetri, Largo E.\ Fermi 5, 50125, 
Firenze, Italy
\and 
ESO, Karl-Schwarzschild-Str. 2, 85748, Garching bei M\"unchen, Germany}

\maketitle              

\begin{abstract}
We are carrying out a search for Ly$\alpha$ emitting galaxies at $z =
6.5$ employing slitless spectroscopy at the VLT. In our 43 arcmin$^2$
field we find three isolated single emission lines with SEDs
consistent with line emitting galaxies at $z = 6.5$.
\end{abstract}

\section{Introduction}

  Since the discovery of $z\sim3$ Lyman Break Galaxies
  (LBGs~\cite{ste96}), many efforts are being made to select the whole
  population of star-forming galaxies at the highest possible
  redshifts (see~\cite{ste99} for a review), and to constrain the star
  formation history of the universe (e.g.~\cite{mad96}). The detection
  of $z > 6$ galaxies allows to study the modes of early galaxy
  formation and the interplay between the first galaxies and the
  intergalactic medium (IGM).  The redshift range $6 < z < 7$ is a
  very intriguing time during cosmic evolution, when hydrogen
  reionization is believed to be basically complete and the IGM starts
  to be polluted with metals \cite{gne00,cia03}.

  Because a significant part of the bolometric luminosity of primeval
  star-forming objects escapes as Ly$\alpha$ emission, especially if
  dust is not yet ubiquitous, it seems profitable to search for the
  redshifted Ly$\alpha$ line of high redshift galaxies. The advent of
  ten meter class telescopes made this approach successful and
  efficient, and both narrow-band imaging and optical
  \emph{serendipitous} spectroscopy found Ly$\alpha$ emitters at $3 <
  z < 6$ \cite{ste99,hu00,mal01}. Thanks to the recently increased
  sensitivity of CCDs at $\lambda > 0.8$\,\umu m, the discovery of $z
  > 6$ galaxies has also become possible. Using narrow band filters
  sensitive to wavelengths in the range $9100 < \lambda < 9250$\,\AA,
  i.e.\ Ly$\alpha$ emission at $z \sim 6.5$ and subsequent
  spectroscopy, two groups have detected the first three galaxies
  known at $z > 6$. The first of these was found due to the strong
  lensing amplification (4.5) by the cluster Abell 370~\cite{hu02}.
  The other two galaxies were found in a survey made with Subaru
  Suprime-Cam instrument which provides a huge field of view (814
  arcmin$^2$~\cite{kod03}).

\section{Observations and analysis}
  We have used a new approach to search for Ly$\alpha$ emitters at
  $z\sim6.5$ based on slitless spectroscopy in combination with a
  narrow band (2.2\%) filter and very deep imaging in a field with
  very low Galactic extinction ($E_{B-V}$=0.002). The $z_{sp}$ filter
  employed has a central wavelength of 9135\,\AA\ and FWHM of
  200\,\AA\ and is therefore sensitive to Ly$\alpha$ at $6.44 < z <
  6.61$. The imaging was carried out through the $z_{sp}$ filter (to
  identify the sources of emission lines detected in the
  spectroscopy) and the Bessel $I$ filter (to determine the $I - z$
  colour or spectral break over the Ly$\alpha$ line). The observations
  were carried out with the FORS2 instrument at the VLT during eight
  nights in October, November 2002 and February 2003. Total exposure
  times were 7.6, 6.33 and 4.0 hours for the slitless spectroscopy,
  $z_{sp}$ and $I$ band direct imaging, respectively. Observing
  conditions were excellent resulting in 0.7 arcsec seeing on the
  resulting combined images.  The field was also imaged earlier in
  Bessel $U$ and Gunn $v$ band.

  Three authors (JK, AC, SSA) have independently searched for emission
  lines on the spectroscopic frame, resulting in a combined list of
  161 emission lines. Each emission line can originate from a
  counterpart on the direct image within a region of about 50$''\times
  1''$.  The emission lines are related to 345 possible counterparts
  in the $z_{sp}$ image.  Four of these have $I - z_{sp} > 1.5$, one
  of which is an [O{\sc III}] emitter at $z = 0.82$.  This leaves three
  counterparts which may be Ly$\alpha$ emitters at $z \sim
  6.5$. Furthermore, among the $\sim 4500$ objects detected by
  SExtractor on the $z_{sp}$ image, there are ten other objects with
  these colours, which are considered LBG candidates at $z > 6$.

\section{Conclusions and future work}

  The number of possible Ly$\alpha$ emitters at $z = 6.5$ detected in
  our field is consistent with the number densities of $z > 5.5$
  galaxies reported by other authors~\cite{ste99,hu00}.  In
  particular, the recent Subaru survey by~\cite{kod03} resulted in the
  detection of 73 Ly$\alpha$ candidates at $z = 6.5$ (with $i' - z' >
  1.3$) in an 814 arcmin$^2$ field.  Based on this result, we expect
  to find in our available 43 arcmin$^2$ field three candidate
  Ly$\alpha$ emitters at $z = 6.5$.

  Slitmask spectroscopic observations are needed to confirm the
  identity of the candidate Ly$\alpha$ emitters and the LBGs.  These
  observations will provide higher signal to noise and higher spectral
  resolution to measure the characteristic asymmetric line profile of
  Ly$\alpha$ at high redshift.  The larger wavelength coverage will
  allow the detection of emission lines from the presumed LBGs outside
  the $z_{sp}$ filter.  Time has been allocated to carry out these
  observations with FORS2.

\end{document}